# Realization of Weyl elastic metamaterials with spin skyrmions

Yuang Pan[1,2,3,4], Liang Si[5], Miao Yang[5], Ning Han[6], Li Zhang[1,2,3,4], Qiaolu Chen[1,2,3,4], Rui Zhao[1,2,3,4], Fujia Chen[1,2,3,4], Yudong Ren[1,2,3,4], Wenhao Li[1,2,3,4], Yuze Hu[1,2,3,4], Mingyu Tong[1,2,3,4], Xinrui Li[1,2,3,4], Junyao Wu[1,2,3,4], Ronghao Bao[5], Weiqiu Chen[5], Yang Long[7,*], Bin Wu[5,*], Hongsheng Chen[1,2,3,4], Baile Zhang[7,8,*], Yihao Yang[1,2,3,4,*]

[1] State Key Laboratory of Extreme Photonics and Instrumentation, ZJU-Hangzhou Global Scientific and Technological Innovation Center, Zhejiang University, Hangzhou 310027, China.

[2] International Joint Innovation Center, The Electromagnetics Academy at Zhejiang University, Zhejiang University, Haining 314400, China

[3] Key Lab. of Advanced Micro/Nano Electronic Devices & Smart Systems of Zhejiang, Jinhua Institute of Zhejiang University, Zhejiang University, Jinhua 321099, China

[4] Shaoxing Institute of Zhejiang University, Zhejiang University, Shaoxing 312000, China

[5] Key Laboratory of Soft Machines and Smart Devices of Zhejiang Province, Soft Matter Research Center, and Department of Engineering Mechanics, Zhejiang University, Hangzhou 310027, China

[6] College of Optical and Electronic Technology, China Jiliang University, 310018 Hangzhou, China

[7] Division of Physics and Applied Physics, School of Physical and Mathematical Sciences, Nanyang Technological University, 21 Nanyang Link, Singapore 637371, Singapore

[8] Centre for Disruptive Photonic Technologies, Nanyang Technological University, Singapore 637371, Singapore

*Correspondence to: (Y.Y.) yangyihao@zju.edu.cn; (B.Z.) blzhang@ntu.edu.sg; (B.W.) bin.wu@zju.edu.cn; (Y.L.) yang.long.physics@outlook.com

# Abstract

Topological elastic metamaterials provide a topologically robust way to manipulate the phononic energy and information beyond the conventional approaches. Among various topological elastic metamaterials, Weyl elastic metamaterials stand out, as they are unique to three dimensions and exhibit numerous intriguing phenomena and potential applications. To date, however, the realization of Weyl elastic metamaterials remains elusive, primarily due to the full-vectorial nature of elastic waves and the complicated couplings between polarizations, leading to complicated and tangled three-dimensional (3D) bandstructures that unfavorable for experimental demonstration. Here, we overcome the challenge and realize an ideal, 3D printed, all-metallic Weyl elastic metamaterial with low dissipation losses. Notably, the elastic spin of the excitations around the Weyl points exhibits skyrmion textures, a topologically stable structure in real space. Utilizing 3D laser vibrometry, we reveal the projection of the Weyl points, the Fermi arcs and the unique spin characteristics of the topological surface states. Our work extends the Weyl metamaterials to elastic waves and paves a topological way to robust manipulation of elastic waves in 3D space.

# Main

Mechanical vibrations in solids, or elastic waves, are ubiquitous, ranging from the solid motion at the macroscopic scale to the lattice oscillations at the nanometer scale. Engineered elastic material, known as elastic metamaterials, can endow elastic waves with topological properties, enabling significant application potential by easing the strict constraints imposed by fabrication tolerances, and establishing a robust framework for routing energy and information[1, 2]. This is crucial for a wide range of applications, including vibration harnessing[3], information processing[4, 5], nondestructive testing[6], and sensing[7]. Particularly, the topological protection in elastic waves can prevent backscattering in high-density phononic circuits, enabling high-efficiency and low-loss phononic circuits with shrunken footprints compared with the photonic counterparts[8], as the phononic wave has much smaller wavelengths than the electromagnetic wave with the same frequency.

Among various topological phases, Weyl semimetal phases are particularly interesting, which, different from topological insulators, are unique to three dimensions and cannot find their counterparts in two dimensions. Metamaterials with Weyl phases exhibit two-fold linear degenerate points of the bandstructure in 3D momentum space[9-14], known as Weyl points, which act as monopoles of Berry flux and carry chiral topological charges measured by the Chern number. Weyl metamaterials can exhibit numerous intriguing phenomena, such as Fermi-arc surface states[15], quantum anomalous Hall effect[16, 17], and anomalous scattering[18, 19], with many promising applications, such as Veselago lenses[18], one-way topological fibre[20], and chiral bulk transport[21].

Though Weyl phases have been widely explored in various physical systems, including electronic, cold-atom, photonic [9, 15, 22-29], and air-borne acoustic systems[12, 14, 30-32], the experimental progress of Weyl elastic metamaterials significantly lags. The difficulty in experimentally exploring the Weyl elastic metamaterials lies in the fact that the elastic waves are intrinsically full-vectorial, with both longitudinal and transverse components, while other waves typically have either purely transverse components (e.g., electromagnetic waves) or purely longitudinal components (e.g., air-borne acoustic waves). On one hand, the full-vectorial nature of elastic waves results in abundant geometrical and spin-related phenomena, such as skyrmion, a topologically stable configuration of a three-component vector field in two dimensions[33, 34]. On the other hand, the full-vectorial nature and the complex couplings between the longitudinal and transverse components of elastic waves lead to complicated and tangled bandstructures, particularly in 3D elastic metamaterials. These bandstructures are highly unfavourable for material design and experimental characterization, as the topological bands tend to coexist with the non-topological counterparts. Moreover, while most 3D air-borne acoustic topological metamaterials made of polymers-like resins have relatively low damping, the 3D polymer elastic metamaterials usually suffer from high damping, owing to the high viscoelasticity of polymers, impeding the experimental demonstration of their topological properties. Thus, a better material choice is highly desired for Weyl elastic metamaterials. Due to these challenges, the experimental realization of Weyl elastic metamaterials has remained elusive to date.

Here, we report the experimental realization of high-quality all-metallic Weyl elastic metamaterials hosting ideal Weyl points with no coexisting non-topological bands. Notably, the elastic spin texture of phononic eigenmodes exhibits topological structures known as spin skyrmions[35], implying the coexistence of real-space and momentum-space topology in our Weyl elastic metamaterials. Importantly, our all-metallic sample has very low vibration damping, and the damping caused by the defects in the sample[36] can be avoided through high-precision 3D metallic additive manufacturing. Using 3D laser vibrometry techniques, we manage to observe the bulk projection of the Weyl points and the topological surface Fermi arcs from the Weyl points over a broad frequency range, and demonstrate the robust propagation and spin characteristics of the topological surface waves.

We design a 3D topological elastic metamaterial that belongs to the space group $P6$ (No. 168), which has six-fold rotational symmetry ($C_6$ symmetry) along the $z$ direction and time-reversal symmetry ($T$ symmetry), as shown in Fig. 1a,b. Each unit cell contains two thin plates connected by two sets of six twisted-placed solid tubes, producing the chiral coupling between the two thin plates. These chiral tubes eliminate all relevant types of rotational and mirror symmetries except $C_6$ symmetry. Lamb waves propagating between the two thin plates couple and interact through these tubes. Employing the first-principles method, the bandstructure of the Weyl elastic metamaterial has been numerically calculated, as displayed in Fig. 1c,f. At the Γ and A points, the first and second bands intersect quadratically in the $k_x$ and $k_y$ directions and linearly in the $k_z$ direction, as do the third and fourth bands. These degenerate points with quadratic dispersion are the so-called charge-2 Weyl points[24]. At the K and H points, the second and third bands intersect linearly in the $k_x$, $k_y$, and $k_z$ directions, which indicates that these degenerate points are charge-1 Weyl points. The $k$-space topology of the Weyl points, i.e., the topological charges of the Weyl points, is defined by the Chern number $C$, which is calculated by integrating the Berry curvature flux on a two-dimensional (2D) manifold **S** enclosing the Weyl point in the momentum space,

$$C = \frac{1}{2\pi} \oint_S \Omega(\boldsymbol{k}) \cdot d\mathbf{S} \tag{1}$$

where $\Omega(\boldsymbol{k})$ is the Berry curvature, $\boldsymbol{k}$ is the wavevector, and $d\mathbf{S}$ is the vector surface element aligned with its local normal direction. By employing the Wilson loop method, we have numerically verified the topological charge of these Weyl points (see Supplementary Note 3).

The Weyl elastic metamaterial and its bandstructure can be explained by a 3D spring-mass model. We take the Wyckoff position $2b$ to construct the lattice, namely two sites in the unit cell. The positions of these two sites are (1/3, 2/3, 0) and (2/3, 1/3, 0), respectively, given by fractional coordinates, which are represented by orange spheres in Fig. 1d,e. The unit cell, depicted in Fig. 1e, has the coupling with chiral force rotation along the $z$ direction, corresponding to the chiral tubes in the Weyl elastic metamaterial. The vibration process is described by two dynamic spring-mass equations (see Supplementary Note 5). Note that contrasting sharply with the previous theoretical proposals for constructing Weyl elastic metamaterials[37-39], which often rely on analogies to existing topological models in electronic systems while overlooking the vectorial features of elastic waves, our approach considers the full-vectorial nature of the vibration modes.

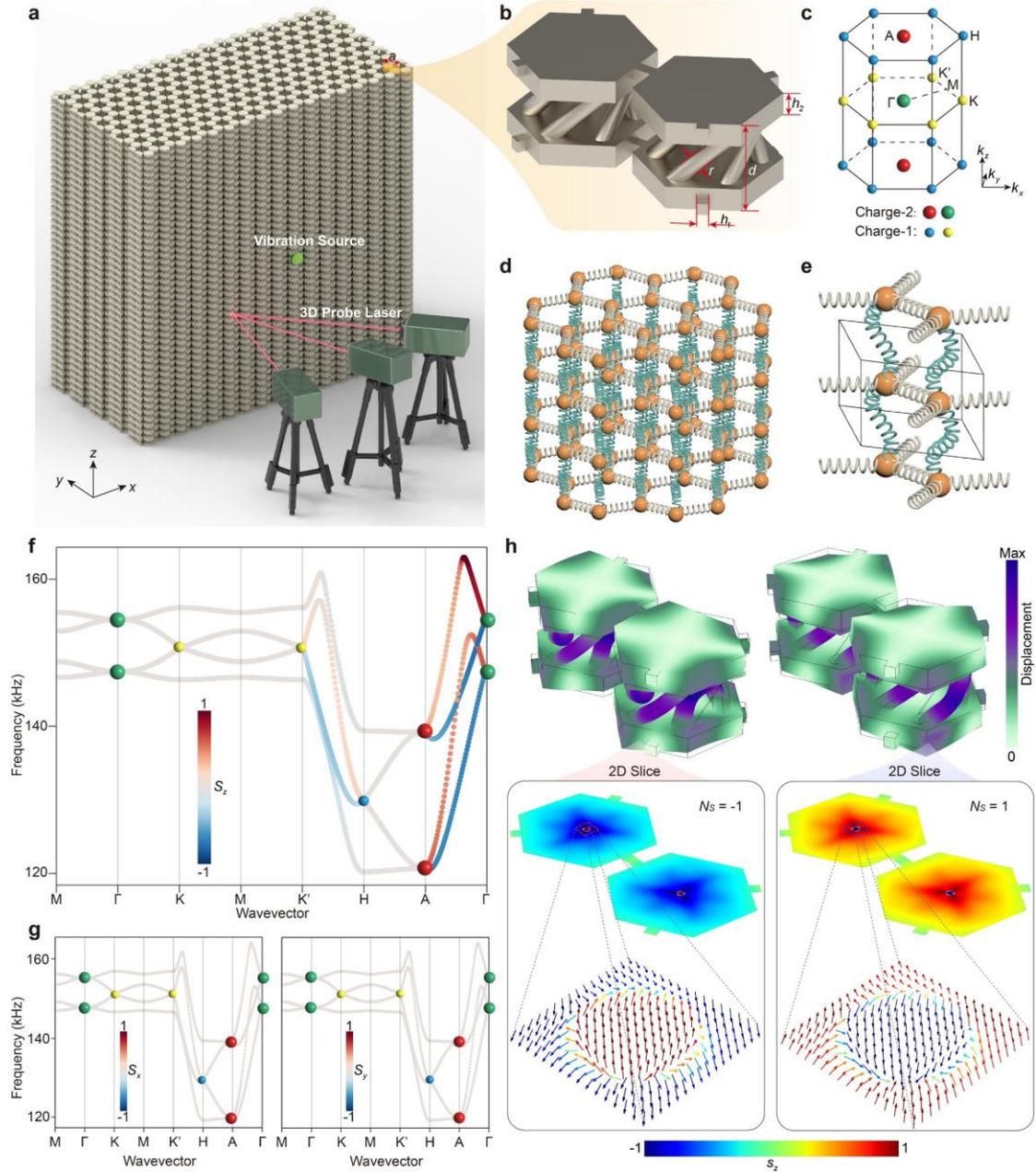

**Fig. 1 Weyl elastic metamaterials with spin skyrmions. a,** Schematic of the experimental setup. The setup consists of a 3D sample, a vibration source, and three laser vibrometers based on enhanced infrared laser technology. **b,** Unit cell of the 3D elastic metamaterial. **c,** 3D Brillouin zone (BZ) of the Weyl elastic metamaterial. The larger spheres represent the charge-2 Weyl points, while the smaller spheres signify the charge-1 Weyl points. **d,** Periodic structure of the spring-mass model. **e,** Zoomed-in view of the spring-mass model. The parallelogram wireframe represents the unit cell. **f,g,** Bulk bandstructure of the Weyl elastic metamaterial, with the colors of the bands denoting the three components of total elastic spin $S_x$, $S_y$, and $S_z$, respectively. **h,** 3D vibration modes and two Néel-type spin skyrmions of the eigenmodes in the first and second bands near Γ.

Interestingly, around the Weyl points at Γ, the eigenmodes exhibit real-space topological structures. The elastic spin characteristic of an eigenmode is described by the elastic spin density $s$[40-43],

$$s = \frac{\rho}{2\omega} \text{Im}\left[ u^* \times u \right] \tag{2}$$

where $\rho$ is the mass density of the material, $\omega$ is the eigenfrequency, and $u(r)$ is the complex displacement of the eigenmodes. The total elastic spin for the eigenmodes, i.e., the integral of the elastic spin density, can be expressed as an integral of $s$, $S = \int_{unit\,cell} s\, dr^3 = (S_x, S_y, S_z)^T$, where $S_x$, $S_y$, and $S_z$ are the three components of the total elastic spin $S$ along the $x$, $y$, and $z$ directions, respectively. As depicted in Fig. 1f,g, under $C_6$ symmetry, the total elastic spin has $S_x = S_y = 0$, and $S_z \neq 0$. However, despite $S_x = S_y = 0$, we show that $s$ can form real-space topological structures--spin skyrmion, which are protected by $C_6$ symmetry[44]. The spin skyrmion is characterized by the skyrmion number[33, 45, 46] $N_s = \frac{1}{4\pi} \int_A m_s \cdot (\partial_x m_s \times \partial_y m_s) dr^2$, where $m_s = s/|s|$ is the unit vector of the spin density, and $A$ denotes a hexagonal cross section of the unit cell. The insets in Fig. 1h show the two spin skyrmions of the eigenmodes in the first and second bands near Γ, both exhibiting Néel-type spin skyrmion textures and carrying oppisite skyrmion numbers $N_s = \pm 1$.

Owing to its self-supporting structure, the 3D elastic metamaterial can be directly fabricated using additive manufacturing techniques. We fabricate a sample from aluminum alloy, consisting of 34×10×44 unit cells, as depicted in Fig. 1a. The specific material of aluminum alloy is $AlSi_{10}Mg$, whose Young's modulus, Poisson's ratio, and density are 70 GPa, 0.33, and 2650 kg/m$^3$, respectively. Note that we initially attempted to choose photosensitive resin as the material for our sample; we, however, figured out that its enormous loss for the propagation of elastic waves prevented us from observing the Weyl points. Consequently, we opt for aluminum alloy to ensure a low loss for elastic waves. The structural parameters are $a$ = 19.5 mm, $r$ = 1.4 mm, $h_1$ = 0.9 mm, $d$ = 7.5 mm, $h_2$ = 1.9 mm.

To extract the 3D vibration characteristics of our Weyl elastic metamaterial sample (see Fig. 2a), we use a 3D laser vibrometer system, as shown in Fig. 1a. The system consists of a source module and a probe module. The source module is a circular lead zirconate titanate (PZT) patch placed on the sample surface to generate elastic waves. The probe module is composed of three laser vibrometers that measure the vibration velocity fields in the $x$, $y$, and $z$ directions simultaneously (the three laser beams are angled, and an internal algorithm decomposes the vibration signal into the $x$, $y$, and $z$ components). To experimentally excite the bulk modes, we place the vibration source three unit-cells away from the back (010) surface, and the bulk vibrations can be observed on the back (010) surface (see Fig. 2b). Then, the surface velocity fields are measured utilizing the 3D laser vibrometers. After applying 2D Fourier transform to the measured field distribution, we obtain the bulk bandstructure projected onto the $k_x$-$k_z$ plane. The results displayed in Fig. 2d reveal that the projected bulk bandstructure along the red line in Fig. 2c hosts a linear band intersection (between the second and third bands) at around 146 kHz. The experimental results are in good agreement with the simulated projected bulk bandstructure shown in Fig. 2e.

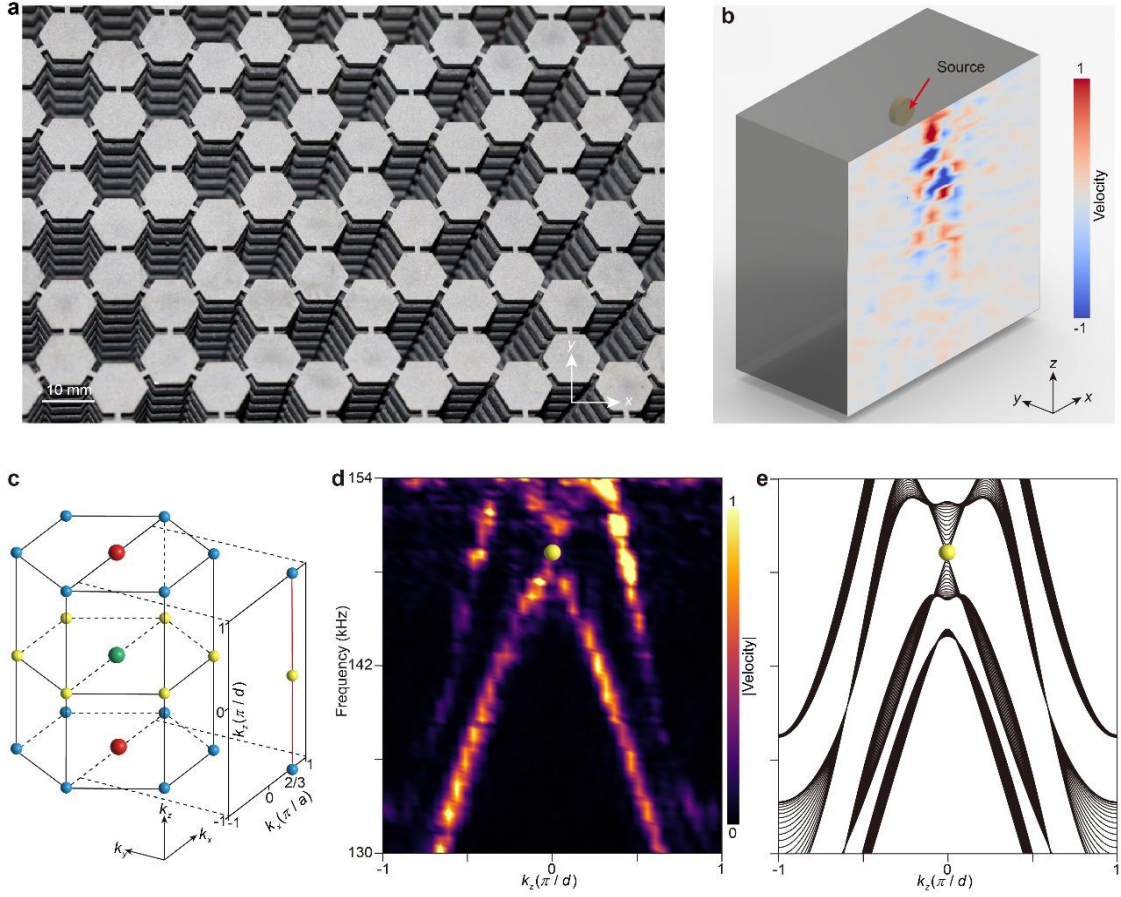

**Fig. 2 Projected bulk bandstructure of the Weyl elastic metamaterial. a**, A detailed photograph of the fabricated sample. **b**, Experimental setup. **c**, The projected 2D surface BZ. **d**, Measured projected bulk bandstructure. **e**, Simulated projected bulk bandstructure.

Next, we perform experiments to measure the topological surface Fermi arcs from the elastic Weyl points on the back (010) surface. As shown in Fig. 3a, a vibration source is placed at the center of the back (010) surface to excite elastic surface waves. Using methods similar to those used for bulk state analysis, we obtain the surface dispersion depicted in Fig. 3b,d,f. One can see that the surface dispersion occupies a broad frequency window spanning from 128 to 152 kHz, with a relative bandwidth of 17.1% (see Fig. 3b), coinciding well with the numerical predictions (see Fig. 3c). Our numerical calculation shows that the displacement of the eigenmodes is well localized on the sample surface (see Fig. 3e). Moreover, we present the measured 2D surface isofrequency contours at frequencies ranging from 130 to 144 kHz in Fig. 3f, which exhibit excellent agreement with the simulated results shown in Fig. 3g. The number of surface Fermi arcs is consistent with the topological charge of the elastic Weyl points. For instance, the two Fermi arcs from the projection of the Weyl point at A confirm that its topological charge is indeed 2.

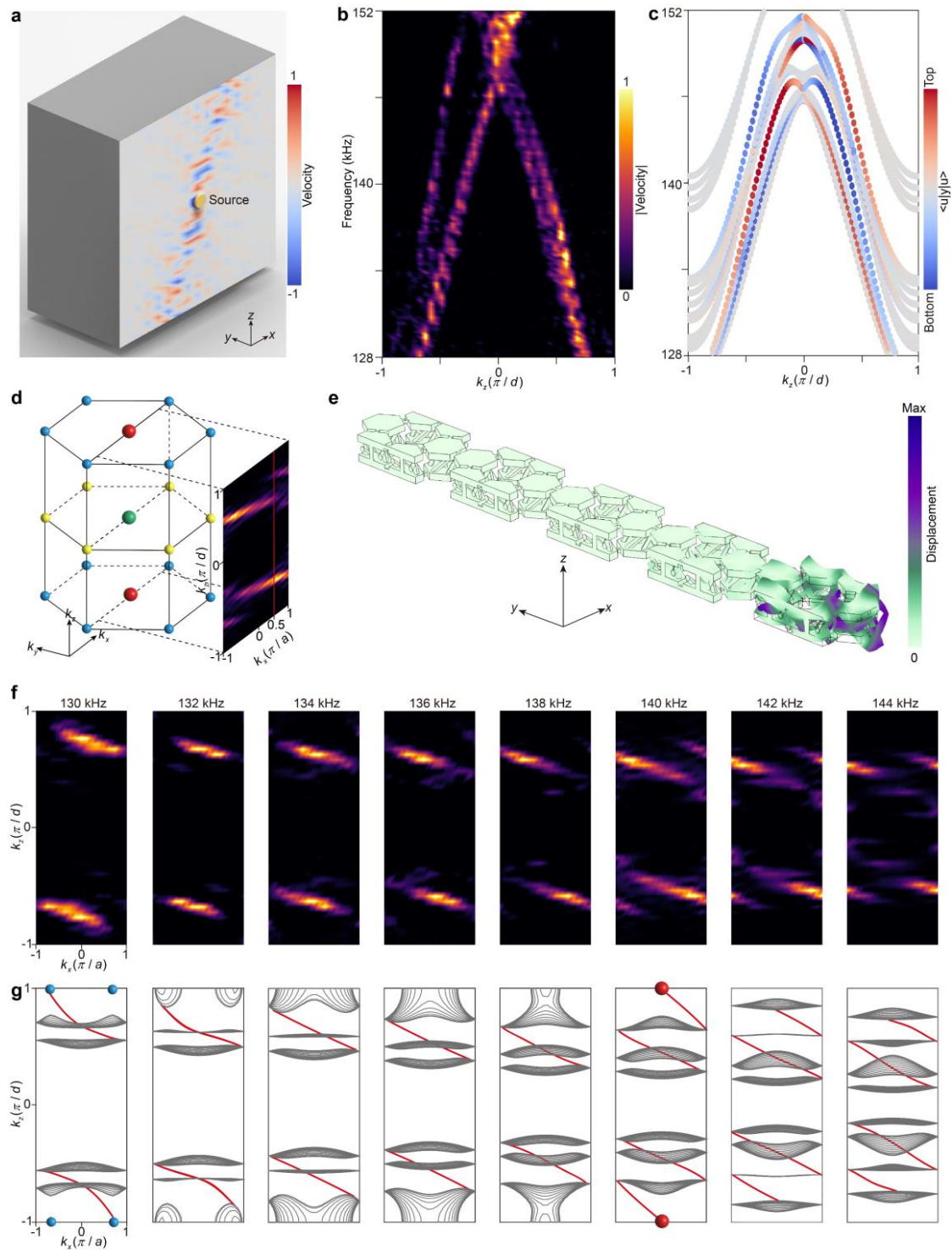

**Fig. 3 Fermi-arc surface states of the Weyl elastic metamaterial. a**, Experimental setup. **b**, Measured surface dispersion with $k_x = \pi/2a$ along the red line in **d**. **c**, Simulated surface dispersion. The color map indicates the corresponding expectation value of position ⟨u|y|u⟩, from which we clearly distinguish the top and bottom surface states. **d**, Schematic of the 2D surface BZ, with the inset colormap displaying the corresponding momentum space results of the field pattern shown in **a**. **e**, Simulated surface vibration mode at 136 kHz. **f**, Measured surface isofrequency contours from 130 to 144 kHz. **g**, Simulated surface isofrequency contours from 130 to 144 kHz.

We also demonstrate the robust 3D propagation of the topological surface elastic waves. As displayed in Fig. 4a, a source is placed at the top of the back (010) surface, and the probe laser scans both the back (010) surface and the back (100) surface. The measured field distribution at 137 kHz, shown in Fig. 4a, clearly demonstrates that the elastic waves can propagate robustly through the corner of the sample without strong backscattering. This is because that for any fixed $k_z$ from -0.7 to -0.5 $\pi/d$, there is only a single $k_x$ value (see Fig. 4b), consistent with our momentum analysis (see Fig. 4c).

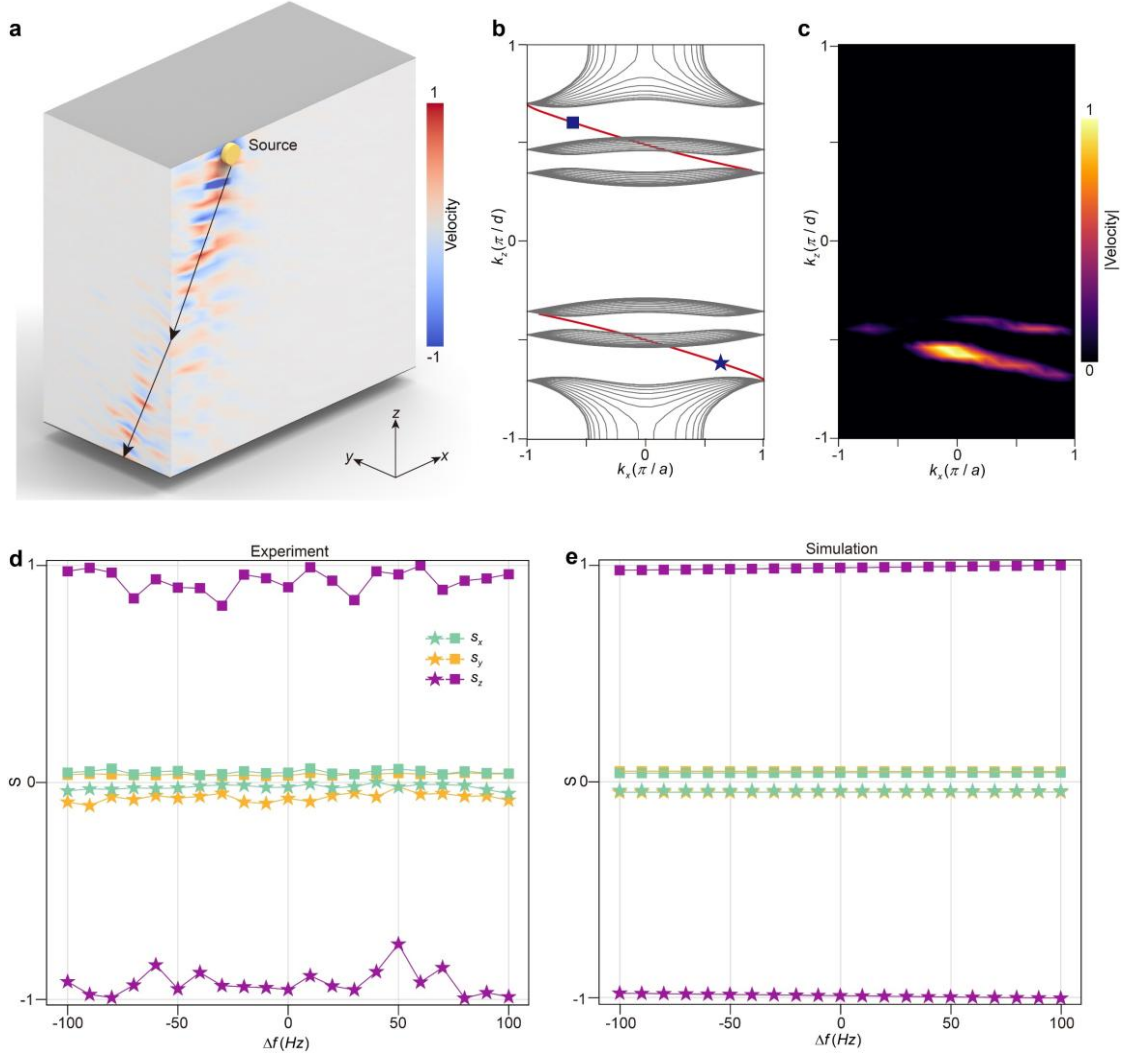

**Fig. 4 Robust 3D propagation and spin characteristics of the elastic Fermi-arc surface waves. a**, Measured field distributions on the back (010) and back (100) surfaces at 137 kHz. **b**, Simulated surface frequency contour at 137 kHz. **c**, 2D Fourier transform of the measured field distribution on the back (010) surface at 137 kHz. **d**, Measured spin density at the two points marked in **b**. **e**, Simulated spin density at the two points marked in **b**. The $\Delta f$ axis represents the frequency deviation from 137 kHz.

Finally, we experimentally investigate the spin characteristics of the topological surface elastic waves. We select two points with opposite $k$ values on the surface Fermi arcs for detection, as displayed in Fig. 4b. The source is placed at the center of the back (010) surface to excite the surface states, and the probe laser is used to detect the two points above and below the source, respectively. Using the measured 3D velocity fields, we calculate the corresponding spin density of the two points around 137 kHz (see Fig. 4d). The results indicate that the topological surface elastic waves propagating in opposite directions possess distinct elastic spin characteristics, i.e., spin-up and spin-down, demonstrating the spin-momentum locking of the surface states.

Our work thus realizes a 3D elastic metamaterial possessing Weyl topology. Through the ingeniously designed 3D topological elastic metamaterial, we have constructed the ideal elastic Weyl points, with the elastic spin of the excitations around the Weyl points exhibiting skyrmion textures. Due to the fully vectorial characteristics of our theoretical model and the high precision and low damping of the fabricated sample, the proposed Weyl elastic metamaterial provides a powerful platform for investigating topological phases in 3D elastic-wave systems, where significant gaps remain to be filled compared to photonic and acoustic systems, including unconventional band degeneracies (e.g., spin-1 Weyl points[19, 47]) and 3D topological insulators with complete and broad bandgaps[48, 49]. For practical applications, the robust and broadband 3D propagation of surface elastic waves in the Weyl elastic metamaterial shows promise for defect-immune vibration waveguiding. The two spin skyrmions with opposite skyrmion numbers in the bulk bands will stimulate potential applications in phononic information encoding and transmission in 3D space. Furthermore, with advancements in 3D printing of all-metallic microstructures enabled by two-photon decomposition technology[50], our design holds potential for on-chip systems, enabling the topological manipulation of phononic waves at the chip level.

## Methods

**Numerical simulations.** All simulations are performed using the COMSOL Multiphysics software package. The Young's modulus, Poisson's ratio, density of the material are set to 70 GPa, 0.33, and 2650 kg/m$^3$, respectively, in alignment with the experimental setup. To calculate the bulk dispersion of the unit cell, periodic boundary conditions are applied in all three spatial directions. To calculate the surface dispersion, a supercell composed of 14 unit cells is considered, with periodic boundary conditions imposed in the *x* and *z* directions, and a free boundary condition in the *y* direction.

**Experiment.** Owing to its self-supporting structure, the sample is fabricated via the metallic additive manufacturing technique. The material is AlSi$_{10}$Mg, with Young's modulus, Poisson's ratio, and density being 70 GPa, 0.33, and 2650 kg/m$^3$, respectively. In the measurements, a circular PZT patch acts as the source to excite the elastic waves. The amplitude and phase of the velocity fields are recorded using the 3D laser vibrometers. To excite the bulk modes, the source is inserted into the sample, and the probe laser scans the surface of the sample. To excite the surface modes, the source is placed at the center of the surface, with the probe laser scanning the corresponding surface.

## Data Availability

All study data are included in the article.

## Acknowledgements

The work at Zhejiang University was sponsored by the Key Research and Development Program of the Ministry of Science and Technology under Grants 2022YFA1405200 (Y.Y.), No. 2022YFA1404704 (H.C.), 2022YFA1404902 (H.C.), and 2022YFA1404900 (Y.Y.), the National Natural Science Foundation of China (NNSFC) under Grants No. 62175215 (Y.Y.), No. 61975176 (H.C.), No. 12402107 (B.W.), No. 12072315 (R.B.), and Nos. 12192210, 12192211 (W.C.), the Key Research and Development Program of Zhejiang Province under Grant No.2022C01036 (H.C.), the Fundamental Research Funds for the Central Universities (2021FZZX001-19) (Y.Y.), the Excellent Young Scientists Fund Program (Overseas) of China (Y.Y. and B.W.), and the China Postdoctoral Innovative Talent Support Program No.BX20240301 (Q.C.). Y.L. gratefully acknowledges the support of the Eric and Wendy Schmidt AI in Science Postdoctoral Fellowship, a Schmidt Futures program.

## Author contributions

Y.Y., B.Z., initiated the idea. Y.Y. and Y.P. designed the experiment, Y.P. and Y.Y. fabricated samples. Y.P. carried out the measurement with the assistance from L.S., M.Y., and R.B.. Y.P. and Y.Y. analysed the data. Y.P., Y.Y., and Q.C. performed the simulations. L.Y., Y.Y., Y.P., L.Z., N.H., R.Z., W.L., X.L., Y.H., M.T., and J.W. did the theoretical analysis. Y.P., Y.Y., L.Y. wrote the paper. Y.Y., H.C., B.Z., B.W., W.C. supervised the project. All authors participated in discussions and reviewed the paper.

## Competing interest

The authors declare no competing interests.

trapping. *Nat. Mater.* 1-9 (2024).